\documentclass[twocolumn]{aastex63}
\usepackage{amsmath}
\usepackage{hyperref}
\usepackage[T1]{fontenc}
\usepackage{xcolor}
\usepackage{soul}
\usepackage{subfigure}
\usepackage{graphicx}
\usepackage{todonotes}
\usepackage[version=4]{mhchem}
\usepackage{isotope}
\usepackage{multirow}
\usepackage{gensymb}
\usepackage{tablefootnote}

\newcommand{\rp}{\emph{r}-process}

\newcommand{\ye}{$\rm{Y_e}$}

\newcommand{\tpr}{$\rm{t_{10}}$}   
\newcommand{\textract}{$\rm{t_{ex}}$} 
\newcommand{\yepr}{$\rm{Y_{e,10}}$}

\newcommand{\bten}{\textsf{b10}}
\newcommand{\bthirty}{\textsf{b30}}
\newcommand{\bcent}{\textsf{b100}}

\newcommand{\nubh}{\textsf{nubhlight}}
\newcommand{\pr}{\textsf{PRISM}}

\begin{document}
\title{Magnetic Field Strength Effects on Nucleosynthesis from Neutron Star Merger Outflows}
\author[0000-0003-0031-1397]{Kelsey A. Lund}
\affiliation{Department of Physics, North Carolina State University, Raleigh, NC 27695 USA}
\affiliation{Center for Nonlinear Studies, Los Alamos National Laboratory, Los Alamos, NM 87545, USA}
\affiliation{Theoretical Division, Los Alamos National Laboratory, Los Alamos, NM 87544 USA}

\author[0000-0001-6811-6657]{Gail C. McLaughlin}
\affiliation{Department of Physics, North Carolina State University, Raleigh, NC 27695 USA}

\author[0000-0001-6432-7860]{Jonah M. Miller}
\affiliation{CCS-2, Los Alamos National Laboratory, Los Alamos, New Mexico 87545, USA}

\author[0000-0002-9950-9688]{Matthew R. Mumpower}
\affiliation{Theoretical Division, Los Alamos National Laboratory, Los Alamos, NM 87544 USA}

\begin{abstract}
Magnetohydrodynamic turbulence drives the central engine of post-merger remnants, potentially powering both a nucleosynthetically active disk wind and the relativistic jet behind a short gamma ray burst.
We explore the impact of the magnetic field on this engine by simulating three post-merger black hole accretion disks using general relativistic magnetohydrodynamics with Monte Carlo neutrino transport, in each case varying the initial magnetic field strength.
We find increasing ejecta masses associated with increasing magnetic field strength.
We find that a fairly robust main \rp{} pattern is produced in all three cases, scaled by the ejected mass.
Changing the initial magnetic field strength has a considerable effect on the geometry of the outflow and hints at complex central engine dynamics influencing lanthanide outflows.
We find that actinide production is especially sensitive to magnetic field strength, with overall actinide mass fraction calculated at 1 Gyr post-merger increasing by more than a factor of six with a tenfold increase in magnetic field strength. This hints at a possible connection to the variability in actinide enhancements exhibited by metal poor, \rp-enhanced stars.
\end{abstract}


\section{Introduction}

The detection of the electromagnetic transient accompanying the binary neutron star merger (NSM) GW170817 lent strong support to these sites as a primary site for the production of the heaviest elements via the rapid neutron capture process (\rp) \citep{Abbott2017,Abbott2017a,Alexander2017,Cowperthwaite2017,Villar2017}. The decay of radioactive species produced through the \rp{} powers the electromagnetic transient that follows the merger event \citep{Lattimer1974,Lattimer1976,Li1998,Metzger2010,Roberts2011,Barnes2013,Grossman2014,Wollaeger2018,Fontes2020}. Interpretation  of the multi-wavelength electromagnetic signal, AT2017gfo, points to at least two components to the ejecta \citep{Chornock2017,Cowperthwaite2017,Nicholl2017,Perego2017}. An early, rapidly decaying signal observed peaking at shorter wavelengths is generally attributed to ejecta with negligible lanthanide abundances \citep{Metzger2010,Roberts2011,Evans2017, Miller2019,Ekanger2023}. Meanwhile, a dim, slowly decaying component observed at longer wavelengths points towards an underlying "red" component generally attributed to a composition of high-opacity, lanthanide-rich ejecta \citep{Barnes2013,Tanaka2013,Kasen2017,Tanvir2017}.

One such site considered to potentially be capable of producing these atomically complex lanthanides is material that is unbound from the accretion disk formed around a remnant black hole \citep{Ruffert1997,Popham1999,Shibata2007,Surman2008,Fernandez2013,Fernandez2014,Janiuk2014,Foucart2015,Just2015,Sekiguchi2015}. As the material is driven off the disk, it is subject to numerous physical processes, each contributing its own uncertainty to the final outcome. One such uncertainty lies in the magnetic fields imprinted on the post-merger system; magnetic fields have long been recognized as playing an essential role in black hole accretion disks \citep{Fernandez2013,Christie2019,deHaas2023}. They influence the dynamics of the accretion process, energy transport, and outflow properties. 

The magnetorotational instability (MRI) of the magnetized plasma \citep{Velikov1959,Balbus1991} leads to the generation of turbulence, enhances angular momentum transport, and is one of the main drivers of the outflow \citep{Shakura1973}. The magnetic field strength can impact the time it takes for this instability to set in, thereby affecting the accretion rate of material onto the black hole as well as the outflow timescales of material off the disk. These outflow timescales are particularly important for the nucleosynthetic yields of the disk when they come into competition with other time scales, especially weak interaction time scales. 

In this work, we aim to investigate the effect of variable initial magnetic field strength on the evolution and nucleosynthetic outcome of the post-merger disk. In Section \ref{sec:method}, we describe the methods we employ to evolve the post-merger disk as well as to carry out nucleosynthesis. We build upon work carried out in \citet{Miller2019} and \citet{Sprouse2023} using three-dimensional, general relativistic neutrino radiation magnetohydrodynamics ($\rm{GR\nu MHD}$) to evolve the disk. In Section \ref{sec:outflows}, we present the results of our simulations and investigate differences in mass outflow as influenced by initial magnetic field strength, as well as the impact of these differences on the conditions in which nucleosynthesis takes place. In Section \ref{sec:nucres}, we show the results of our nucleosynthesis calculations and investigate the contributions to the total abundances from different spatial components of the mass outflow. Finally, in section \ref{sec:conclusion}, we interpret the broader implications of our results and provide some concluding remarks. 

\section{Method}\label{sec:method}
We seek to quantify the effect of variable initial magnetic field strength on the mass ejection as well as the nucleosynthesis of heavy elements that occurs in the black hole accretion disk formed after a neutron star merger.

\subsection{Post-Merger Disk Evolution}
\begin{figure}[ht]
    \centering
    \includegraphics[scale=0.75]{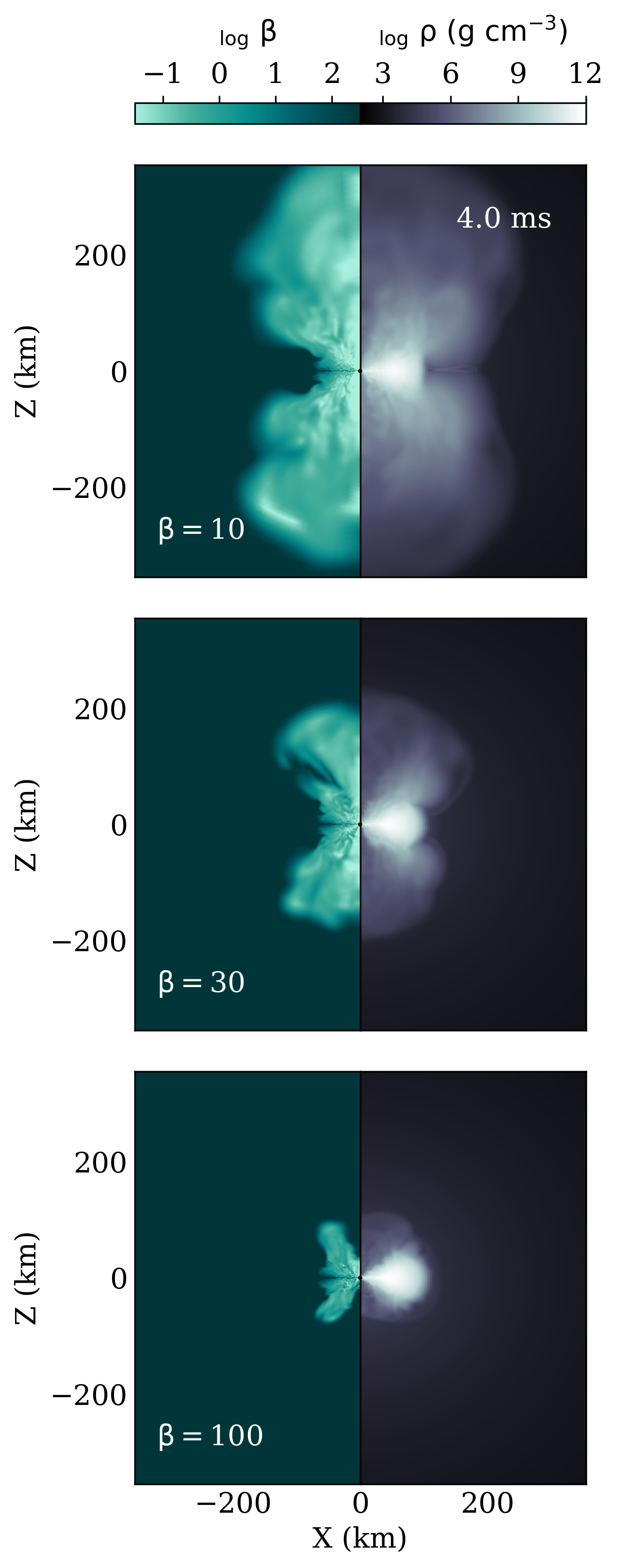}
    \caption{The value of $\beta$ (left) and density (right) for the three different disks with decreasing field strength from top to bottom. The disks are shown at 4 ms physical time post-merger.}
    \label{fig:4ms}
\end{figure}

We use \nubh, which solves the equations of general relativistic ideal magnetohydrodynamics (GRMHD) with neutrino radiation transport, to evolve the post-merger disk. \nubh{} builds on a long history of methods spanning almost two decades \citep{Gammie2003,Dolence2009,Ryan2015,Porth2019}. The methods used in \nubh{} are detailed in \citet{Miller2019_nubh}; we include a summary here for convenience and context.

\nubh{} solves the equations of ideal GRMHD via finite volume methods with constrained transport, and uses Monte Carlo methods to perform neutrino radiation transport. The two are coupled via first-order operator splitting. We use a radially logarithmic quasi-spherical grid in horizon penetrating coordinates, as first described in \citet{McKinney2004}, the WENO reconstruction first described in \citet{Tchekhovskoy2007}, the primitive variable recovery scheme described in \citet{Mignone2007}, and the drift-frame artificial atmosphere treatment described in \citet{Ressler2015}. 

For microphysical data, we use the SFHo equation of state, tabulated in Stellar Collapse format \citep{OConnor2010,OConnor2010_web} and described in \citet{Steiner2013}. For neutrino opacities and cross sections, we use the charged and neutral current interactions as tabulated in \citet{Skinner2019} and described in \citet{Burrows2006}. Our neutrino scattering implementation uses a biasing technique to ensure all processes are well sampled, as described in \citet{Miller2019_nubh}.

We construct three separate \citet{Fishbone1976} torii, representing post-merger disks, each with a single poloidal magnetic field line, the strength of which we vary by varying the dimensionless parameter, $\beta$:
\begin{equation}
    \beta = \frac{\rm{P_{gas}}}{\rm{P_{magnetic}}},
\end{equation}
using values of 10, 30, and 100. Throughout the text, we refer to these as \bten, \bthirty, and \bcent, respectively, and highlight that a smaller value of $\beta$ corresponds to a stronger initial magnetic field. 

In all three cases, we begin with uniform electron fraction, $\rm{Y_e}=0.1$, throughout the disk. We begin with an accretion disk surrounding a $2.58 M_\odot$ \citet{Kerr1963} black hole with dimensionless spin parameter $a=0.69$, as in \citet{Miller2019}, and a disk mass of $0.12 M_\odot$. We use a radially logarithmic grid of dimensions $\rm{N_r}\ x\ \rm{N_\theta}\ x\ \rm{N_\phi} = 192\ x\ 128\ x\ 66$ and run the simulation out to $10^4\ GM_{BH}/c^3$, which corresponds to 127 ms of physical time given this geometry. At any given time, there are roughly $3\times 10^7$ Monte Carlo packets in the radiatively active region of the simulation.

A post-merger disk simulation must be sufficiently resolved in both the grid and Monte Carlo particles to both capture the MRI and the neutrino-matter interactions. Resolution requirements depend on the fastest-growing wavelength of the MRI and the time scale of weak interactions. Broadly, resolution requirements become more severe for weaker initial magnetic fields (and for non-poloidal field topologies). Our systems are all as resolved, or better, than the model presented in \cite{Miller2019}.

In Figure \ref{fig:4ms}, we show each of the disks colored by the value of $\beta$ on the left and density on the right, as they appear 4 ms into each simulation, showing the more rapid time evolution exhibited by the stronger magnetic field.

\subsubsection{Tracers\label{sec:tracers}}
We uniformly sample \citep[by volume, see][]{Bovard2017b} approximately $1.5 \times 10^6$ Lagrangian fluid packets (tracer particles) at the beginning of each simulation everywhere there is physical fluid, i.e. not artificial atmosphere. These tracers are passively advected with the fluid throughout the simulation and provide the thermodynamic evolution that is used to calculation the nucleosynthesis (see section \ref{sec:nucleo}).

Of these tracers, we select those that have reached a radius of at least 250 gravitational radii (corresponding to approximately $10^3$ km in this geometry) by the end of the simulation, and which furthermore have a Bernoulli parameter $B>0$. This ensured that the material we captured was unbound by the end of the \nubh{} simulation. This selection process resulted in 195288, 135901, and 74643 tracers for the \bten, \bthirty, and \bcent{} disks, respectively. When referring to the time at which an individual tracer crosses this boundary, we use the notation \textract. When referring to properties recorded at this same time, we use the same subscript, e.g. $\rm{\theta_{ex}}$.

At some point in the evolution of the disk, each of the tracers in our sample falls below a temperature of 10 GK for the last time, at which point we record their properties as a starting point for nucleosynthesis. We label this time \tpr{}, and any associated tracer property recorded at this time with the same subscript. We emphasize that both the values of \textract{} and \tpr{} can be different for each individual tracer. We further emphasize that both \textract{} and \tpr{} refer to times after the start of the simulation.

\subsection{Nucleosynthesis\label{sec:nucleo}}
We compute abundances for each of the \nubh{} tracers out to 1 Gyr post-merger.

\subsubsection{Thermodynamic Evolution}
We use the portable routines for integrated nucleosynthesis modeling (\pr{}) \citep{Sprouse2021} to perform nucleosynthesis calculations using the trajectories of the tracer particles that emerge from the selection of tracers described in the previous section. For each tracer, beginning at \tpr, we extract the temperature and density as inputs for \pr{}. Given that our \nubh{} calculations are only run out to 127 ms, we extrapolate the tracers assuming homologous expansion past 127 ms out to at least 1000 seconds. 

We further use \yepr{} together with the SFHo equation of state \citep{Steiner2013} to determine the initial nuclear abundances.

\subsubsection{Nuclear Model Set for Nucleosynthesis}
\pr{} is designed to take input files describing the decays and reactions to which nuclei involved in the r-process are subjected. We use the JINA Reaclib library \citep{Cyburt2010} for charged-particle and light-nuclei interactions. We utilize the theoretical beta decay rates from \citet{Moller2019} and compute both beta-delayed neutron emission as well as beta-delayed fission probabilities using \citet{Mumpower2016}.

We compute neutron capture and neutron-induced fission rates using the statistical Hauser-Feshbach code CoH \citep{Kawano2016}. We use the barrier height-dependent prescription from \citet{Karpov2012} and \citet{Zagrebaev2011} to calculate spontaneous fission rates using the FRLDM barrier height description \citep{Moller2015}. Our theoretical alpha decay rates computed using the Viola-Seaborg relation. Finally, where experimental or evaluated data exists, we overwrite theoretical values with data from the 2020 version of NuBase \citep{Kondev2021} and the Atomic Mass Evaluation \citep{Wang2021}. 

\subsubsection{Angle Dependence}\label{sec:angle}
We are interested in identifying and distinguishing material that is ejected at small angles above the equator (equatorial material) from material that is ejected at larger angles above the equator. In particular, the formation of a relativistic jet can eject material on very short timescales. In contrast, viscously driven material remains in the disk and evolves on slower timescales. Additionally there may be a thermally-driven wind that can launch outflows on higher-latitude trajectories more rapidly than viscous driving \citep{Fernandez2013, Siegel2017,Fernandez2018,Miller2019,Fahlman2022,Kiuchi2023,Sprouse2023}.

These differences in timescales can result in vastly different conditions for the material as it begins to undergo nucleosynthesis. Traditionally, the faster evolving polar material tends to be associated with very poor lanthanide production and almost no actinide production. On the other hand, the equatorial, viscous material tends to be associated with a higher opacity due to a more lanthanide-rich composition. 

To probe these different regimes, we define criteria for classifying individual tracers as either part of the equatorial viscous bulk, part of the polar bulk of material, or something in between. We classify a tracer as being polar if its spatial position at \textract{} lies 45 degrees or more \emph{above} the mid-plane. Alternatively, we classify a tracer as being equatorial if its spatial position at \textract{} lies 15 degrees or \emph{less} above the mid-plane. All tracers which do not satisfy either of these criteria fall into the "intermediate" category. 

\section{Outflow Properties}\label{sec:outflows}

We describe the properties of the material as it begins to undergo nucleosynthesis, as well as its properties as it becomes unbound from the system.

\subsection{Mass Ejection}
\begin{figure}[ht]
    \centering
    \includegraphics[scale=0.4]{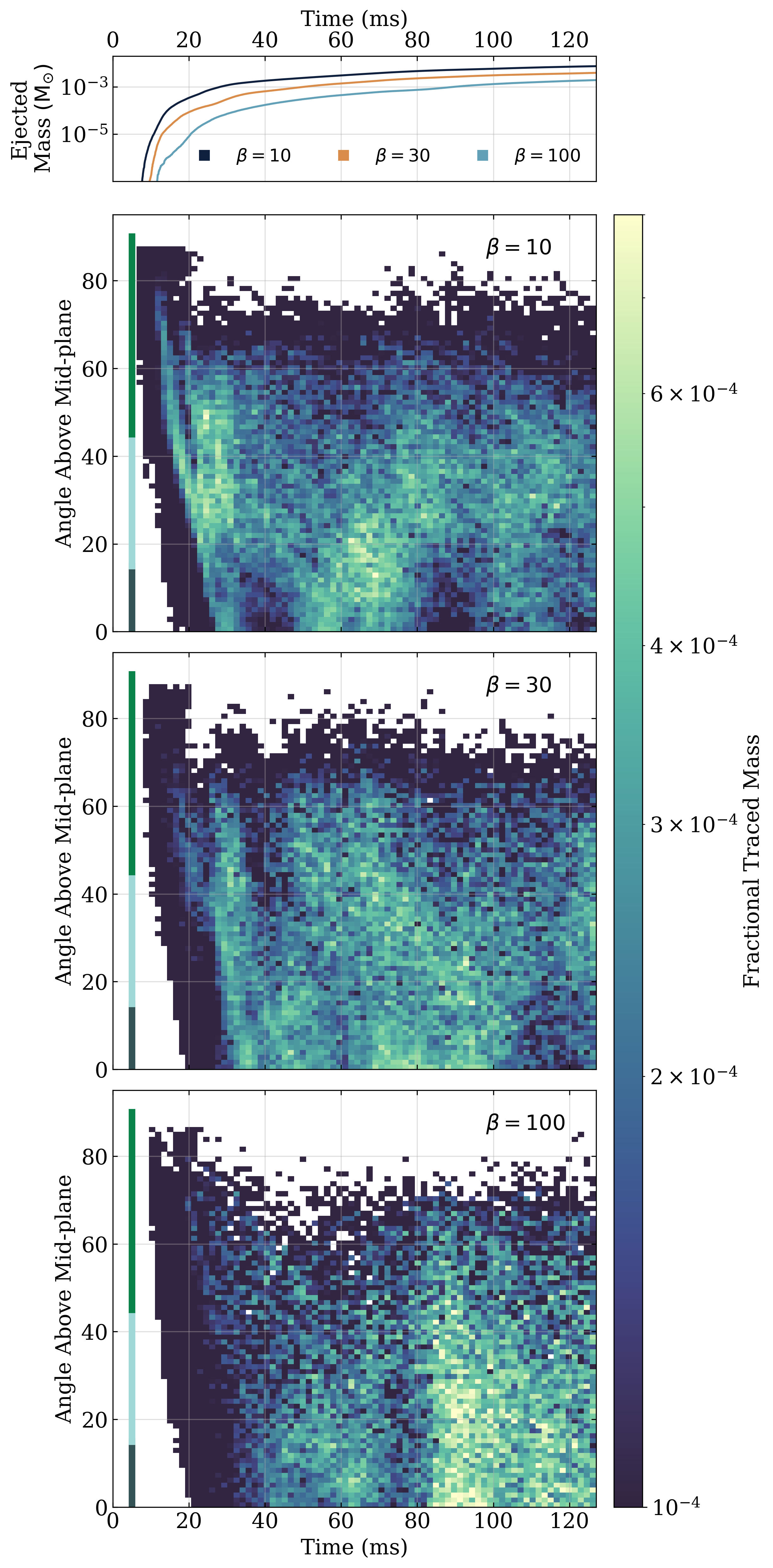}
    \caption{\emph{Top panel:} Cumulative mass ejected for each disk as a function of time. \emph{Bottom panels:} Relative (fractional) mass ejected through $\rm{r_{ex}}$ as a function of time and angle above the mid-plane for each disk outflow. In each case, the values of the histograms sum to one. }
    \label{fig:massoutflow}
\end{figure}

We record the properties of the Lagrangian tracers as they pass through a sphere of 250 gravitational radii ($\rm{r_{ex}}$, corresponding to roughly $10^3$ km. in this geometry). By the end of the simulation, the total mass that we consider to be unbound is the mass that has passed through this surface. We found that the total mass ejected was largest for the \bten{} disk, and decreased as the relative field strength decreased, as can be seen by the end point of the curves in the top panel of Figure \ref{fig:massoutflow}.

The total ejecta masses from each disk are listed in the top-most row of Table \ref{tab:masses}. By the end of the \bcent{} simulation, the total mass ejected was roughly $2\times10^{-3}\ \rm{M_\odot}$. Decreasing $\beta$ to 30 increased the total ejected mass to $\sim4\times10^{-3}\ \rm{M_\odot}$; decreasing $\beta$ to 10 further increased the ejected mass by almost a factor of 2, unbinding the largest mass of $7\times10^{-3}\ \rm{M_\odot}$.

The bottom three panels of Figure \ref{fig:massoutflow} show the time evolution of the angular distribution of material as it passes through $\rm{r_{ex}}$, with increasing $\beta$ (and therefore decreasing field strength) from top to bottom. We include vertical lines showing the angular cuts described in section \ref{sec:angle} as a guide to the eye. Increasing the magnetic field strength results in a more quickly evolving system. This can also be seen in Figure \ref{fig:4ms}, where the \bten{} disk is further along in its evolution by 4 ms compared to the \bcent{} disk at the same time. 

All three disks showed an early-time (by 20 ms) outflow at elevations above $20\degree$. However, it was the \bten{} disk which showed a large burst of material escaping the system early on- the \bcent{} disk only saw a trace amount of material escape in this time. Specifically, the \bten{} disk had ejected more than 4\% of the total mass ejected by the time the simulation reached 20 ms. For comparison, the \bthirty{} disk at the same time had ejected roughly 2\% of the total ejected mass, and the \bcent{} disk only about 0.4\%.

The disks with higher magnetic field tended to evolve more quickly and therefore ejected a greater proportion of their mass more quickly.  Just after the halfway point of the simulation, around 67 ms, the \bten{} disk has already ejected half its material, with \bthirty{} following closely at 70 ms and \bcent{} at 87 ms. This trend towards earlier overall mass ejection as a function of increasing field strength represents at least two effects. The first is the relative mass ejected in this early ($<20$ ms) transient while the second is the behavior of hot spots of increased mass ejection, such as can be seen after about 80 ms in the \bcent{} disk in Figure \ref{fig:massoutflow}. 

We note that these values represent lower limits to the total mass ejected from the system. The top panel of Figure \ref{fig:massoutflow} shows a cumulative mass ejection curve that has not reached a plateau by the end of the simulation, i.e. the ejected mass would continue to increase. We direct the reader to \citet{Sprouse2023}, for a recent study on the \bcent{} disk evolved a factor of ten longer, and the impacts on mass ejection and nucleosynthesis.

\begin{table*}
    \centering
    \begin{tabular}{|c | ccc | ccc | ccc|}
        \hline
         &  &  &  &  &  &  &  &  & \\
         & \multicolumn{3}{c|}{\bten} & \multicolumn{3}{c|}{\bthirty} & \multicolumn{3}{c|}{\bcent}\\
         \hline
         Total Mass ($\rm{M_\odot}$) & \multicolumn{3}{c|}{$7.31\times10^{-3}$} & \multicolumn{3}{c|}{$3.83\times10^{-3}$} & \multicolumn{3}{c|}{$1.89\times10^{-3}$}\\
         $\rm{M_{Lan}\ (M_\odot)}$ & \multicolumn{3}{c|}{$2.79\times10^{-4}$} & \multicolumn{3}{c|}{$1.0\times10^{-4}$} & \multicolumn{3}{c|}{$6.65\times10^{-5}$}\\
         $\rm{X_{Lan}\ (\times10^{-2})}$ & \multicolumn{3}{c|}{3.82} & \multicolumn{3}{c|}{2.62} & \multicolumn{3}{c|}{3.52}\\
         $\rm{M_{Act}\ (M_\odot)}$ & \multicolumn{3}{c|}{$5.03\times10^{-5}$} & \multicolumn{3}{c|}{$7.84\times10^{-6}$} & \multicolumn{3}{c|}{$2.07\times10^{-6}$}\\
         $\rm{X_{Act}\ (\times10^{-3})}$ & \multicolumn{3}{c|}{6.88} & \multicolumn{3}{c|}{2.05} & \multicolumn{3}{c|}{1.1}\\
         $\rm{M_{Lan}/M_{Act}}$ & \multicolumn{3}{c|}{5.5} & \multicolumn{3}{c|}{12.8} & \multicolumn{3}{c|}{32.2}\\
         \hline
         \hline
         \hline
         & \multicolumn{3}{c|}{Equatorial} & \multicolumn{3}{c|}{Intermediate} & \multicolumn{3}{c|}{Polar}  \\
         & \bten & \bthirty & \bcent & \bten & \bthirty & \bcent & \bten & \bthirty & \bcent\\
         \hline
         Percent of Total Ejecta Mass & 19.8 & 23.5 & 27.5 & 53.8 & 48.3 & 48.8 & 26.4 & 28.2 & 23.8\\
         Percent of Total Lanthanide Mass & 50.8 & 64.8 & 48.8 & 41.7 & 23.8 & 48.8 & 7.46 & 11.4 & 2.4\\
         Percent of Total Actinide Mass  & 62.6 & 68.6 & 36.7 & 30.1 & 13.5 & 60.6 & 7.34 & 17.9 & 2.80\\
         \hline
         $\rm{X_{Lan}\ (\times10^{-2})}$ & $9.79$ & $7.24$ & $6.26$ & $2.96$ & $1.29$ & $3.53$ & $1.08$ & $1.06$ & $0.353$\\
         $\rm{X_{Act}\ (\times10^{-3})}$ & $21.7$ & $5.98$ & $1.46$ & $3.84$ & $0.573$ & $1.36$ & $1.91$ & $1.30$ & $0.129$\\
         \hline
         $\rm{M_{Lan}/M_{Act}}$ & 4.5 & 12.1 & 42.9 & 7.7 & 22.5 & 26.0 & 5.6 & 8.2 & 27.3\\
         \hline
    \end{tabular}
    \caption{\emph{Top:} Global outflow properties for each disk. \emph{Bottom:} Mass and composition properties for each angular cut within each disk.}
    \label{tab:masses}
\end{table*}

\subsection{Nucleosynthetic Conditions}

\begin{figure}[ht]
\centering
\includegraphics[scale=0.4]{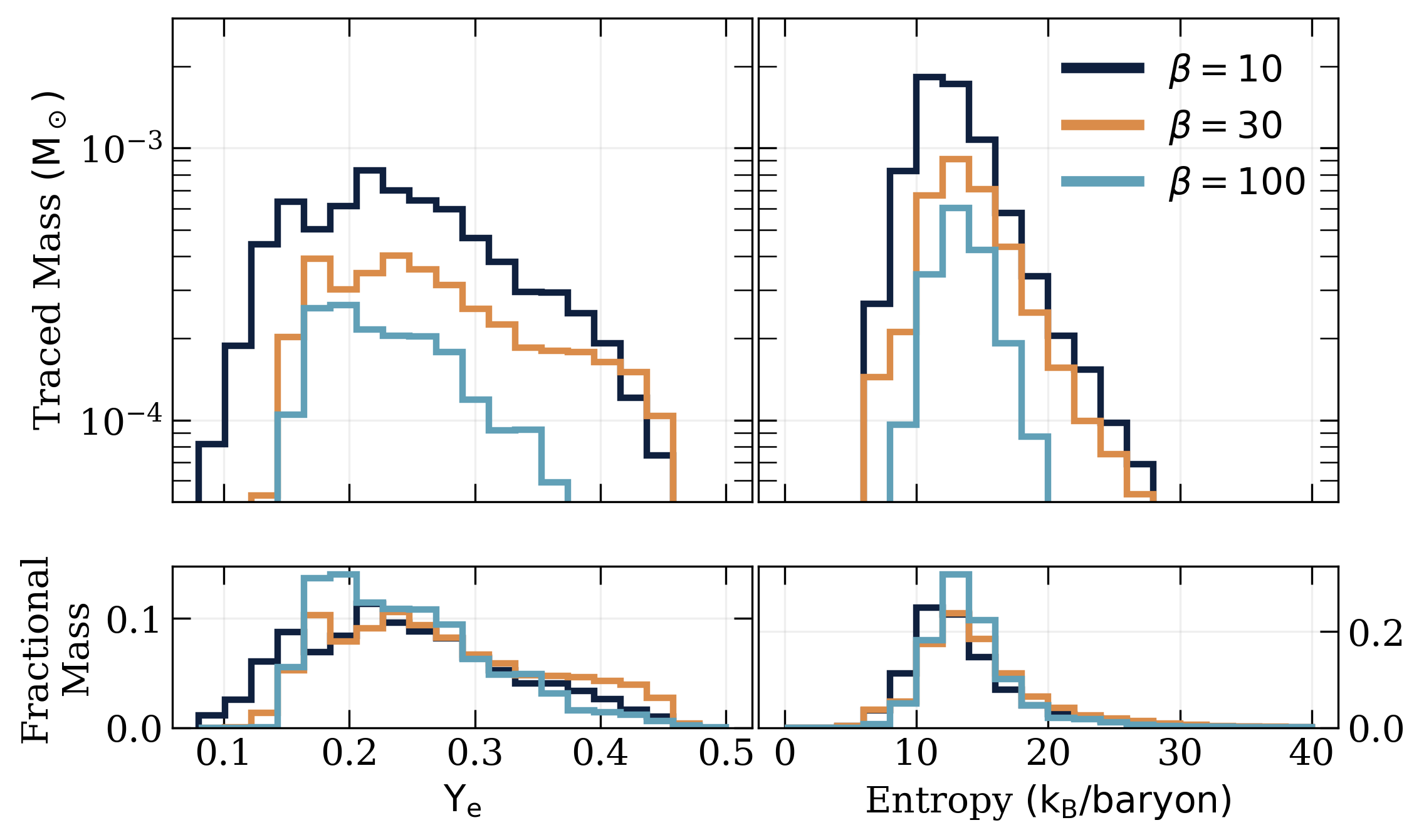}
\caption{\emph{Top row:} Distribution of electron fraction (left), and entropy (right) at \tpr, showing total mass in each bin in units of $\rm{M_\odot}$. \emph{Bottom row:} The same as the top row but showing the fractional mass distribution.}
\label{fig:basics}
\end{figure}

The values recorded at \tpr{} set the conditions for nucleosynthesis. We show the electron fraction (\yepr{}) and entropy ($\rm{s_{10}}$) recorded at \tpr{} in Figure \ref{fig:basics}. The top row of the figure shows the absolute mass distribution, i.e. in units of $\rm{M_\odot}$. The bottom row shows the fractional mass distribution, thereby showing the relative spread in values of \ye{} and entropy.

Given that the electron fraction determines the number of free neutrons available for capture, it is the property most readily associated with producing a more robust \rp{} pattern. A key quantity of interest is the amount of material with $\rm{Y_{e,10} \lesssim 0.25}$ given that this is a rough threshold below which a full \rp{} pattern is considered accessible, although the exact \ye{} threshold  varies according to the entropy and outflow timescale of the material. For all three disks, the tracers with $\rm{Y_{e,10}<0.25}$ accounted for more than half the total ejecta, with values of roughly 63\%, 54\%, and 66\% for $\beta\ \rm{=10,\ 30,\ and\ 100}$, respectively.

Material that begins nucleosynthesis with a \ye{} above 0.3 is very unlikely to produce a full \rp. Because it does not have the high opacity associated with actinides and lanthanides, this ejecta on its own is bluer and more quickly evolving than the lanthanide/actinide-rich outflows. Additionally, if material of this type is mixed in with ejecta that \textit{does} produce lanthanides or actinides, it could serve to dilute this outflow, leading to a bluer component. The fraction of higher \ye{} outflow ranged from 31\% for the highest magnetic field, \bten{} disk to 18\% for the lowest magnetic field, \bcent{} disk. As can be seen in the left column of Figure \ref{fig:basics}, the \bten{} disk produces the widest range of initial electron fractions, spanning the range from 0.1 to 0.45.  This is consistent with it producing both the largest mass fraction of low \ye{} material \emph{and} a large mass fraction of high \ye{} material.

\subsubsection{Nucleosynthetic Conditions by Angle}

\begin{figure}[ht]
\centering
\includegraphics[scale=0.37]{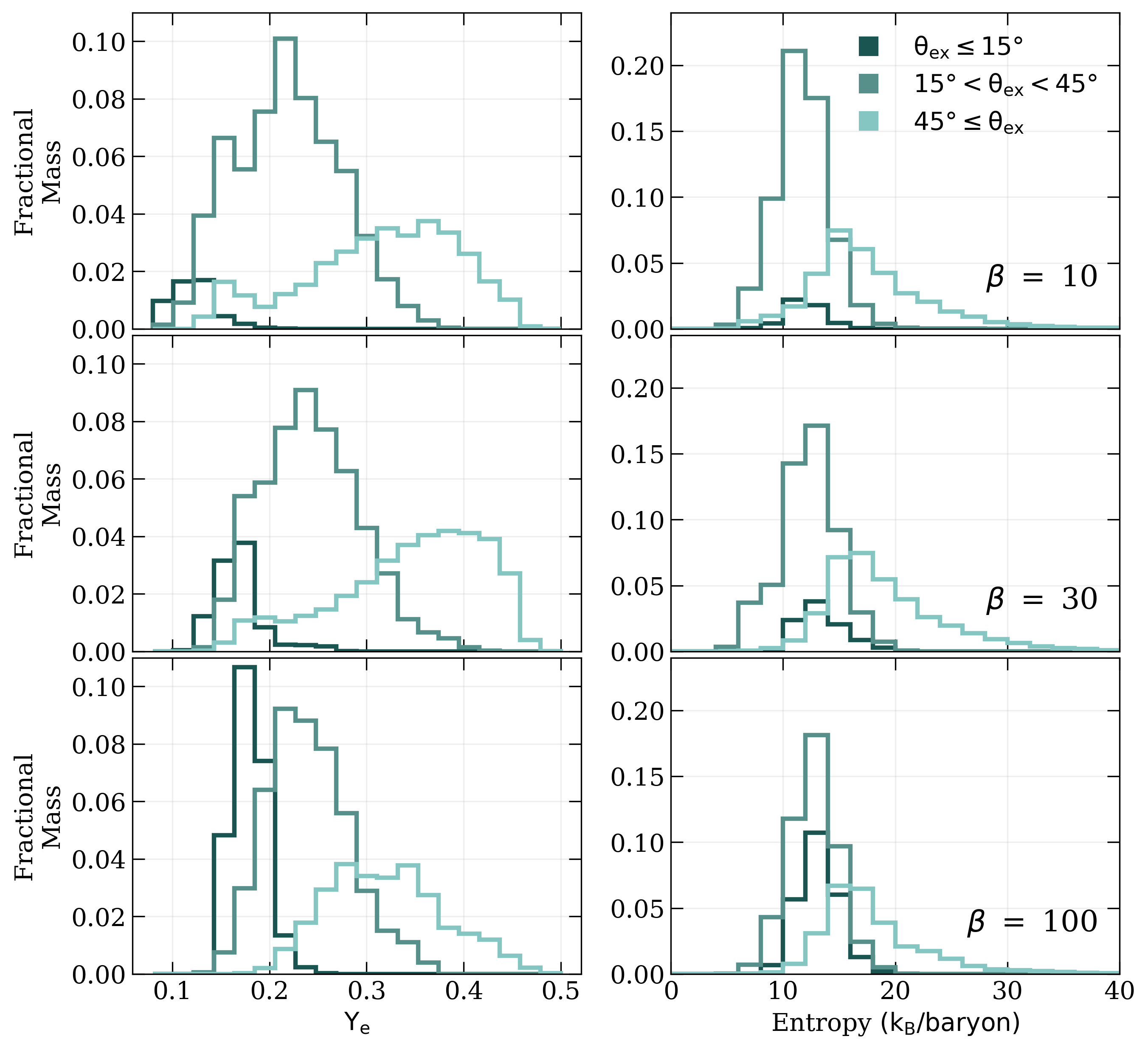}
\caption{Fractional mass distributions of key properties divided into angular cuts for \bten{} (top row), \bthirty{} (middle row), and \bcent{} (bottom row). The left column shows electron fraction while the right column shows entropy.}
\label{fig:basics_angle}
\end{figure}

More subtle differences in the conditions for nucleosynthesis emerge when we investigate these conditions as a function of angle. Figure \ref{fig:basics_angle} shows similar information as Figure \ref{fig:basics}, but in this figure our results are separated into the angular cuts described in section \ref{sec:angle}. 

For all three disks (shown in ascending order of $\beta$ in the three rows of Figure \ref{fig:basics_angle}), the lowest \ye{} material is ejected from the equatorial ($\rm{r_{ex} \leq 15}$) region, with the majority of the equatorial material falling into a range of \yepr{} values at or below 0.25. However, the bulk of the material tends to sit at slightly higher \ye{} with decreasing magnetic field, i.e.  increasing $\beta$.  At polar angles, the typical electron fraction is highest with the bulk of the material being between 0.3 and 0.4, and a tailing off of the distribution above $\rm{Y_e} \sim 0.4$.  Both the \bten{} and \bthirty{} polar ejecta show a sub-dominant peak on the low \ye{} end of the distribution. The intermediate angle material, which tends to make up the majority of the outflow in each disk also has intermediate \ye{} values. The \ye{} distribution of the intermediate-angle material shows less of a dependence on $\beta$, although the \bten{} material shows a sub-dominant low-\ye{} peak around 0.15. 

\section{Nucleosynthesis Results}\label{sec:nucres}

\begin{figure}[ht]
    \centering
    \includegraphics[scale=0.38]{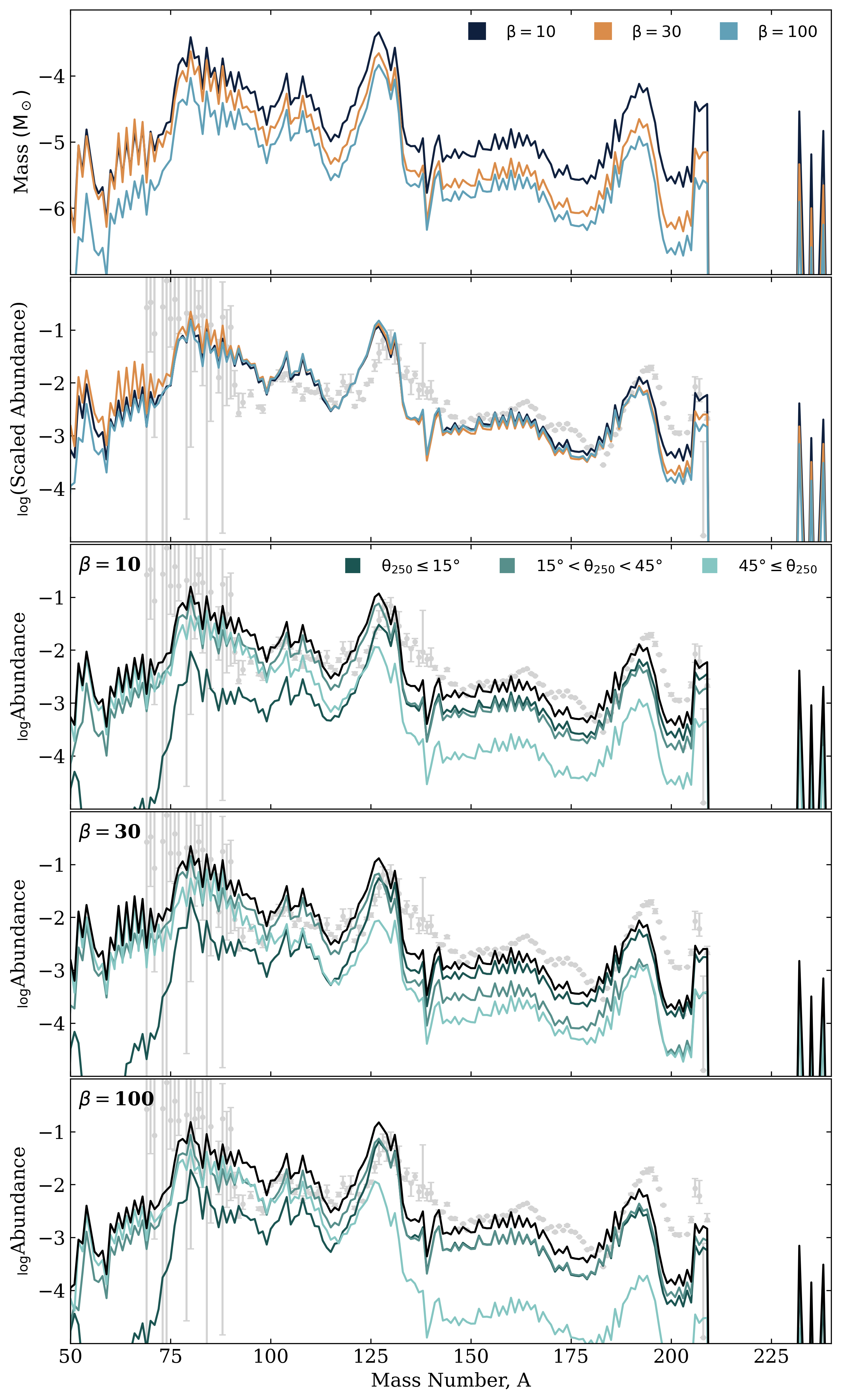}
    \caption{\emph{Top panel:} Isotopic mass for all three disk outflows. \textit{Second panel:} Abundances for all three disk outflows, compared to solar (grey points). All abundances are scaled to the value of A=120 from \bten. \emph{Bottom panels:} Total abundance for \bten{} (third panel), \bthirty{} (fourth panel), \bcent{} (bottom panel) outflows as well as the contribution from the material in each angular cut. }
    \label{fig:abs_angle}
\end{figure}

\subsection{Global Results}

We compare the overall nucleosynthetic outcomes for all tracers from the \bten, \bthirty, and \bcent{} disks. 
We discuss both the overall results as well as the behavior of the cuts described in section \ref{sec:angle}.

The top-most panel of Figure \ref{fig:abs_angle} shows the isotopic masses produced by the tracers from the \bten{} (dark blue), \bthirty{} (orange), and \bcent{} (light blue) disks, while the second panel shows the overall, mass-weighted abundances produced by the tracers from each disk in the same colors. These scaled abundances are scaled to the abundance of A=120 in \bten, and the scaled solar abundances \citep{Arlandini1999,Goriely1999} are shown as grey points. We find broad agreement in the overall shape of the abundance patterns, though with notable differences in the scale of the patterns and in particular regions.

By scaling the abundances produced in each disk, we are able to compare the relative abundances in different regions to $\rm{A=120}$. As noted in the last section, the highest magnetic field disk (\bten), has the largest fraction of low \ye{} material. As expected, the outflow from this disk is most effective at producing lanthanide, third peak and actinide material relative to the second peak. One might similarly expect that the \bten{} disk is also most effective at producing the relative amounts of light r-process elements but in fact, it is the intermediate magnetic field disk that does this. This is consistent with Figure \ref{fig:basics} where the \bthirty{} disk has the largest fractional ejected mass at high \ye. 

The top section of Table \ref{tab:masses} lists the total mass unbound for each of the three simulations, as well as what mass fraction of that outflow was composed of lanthanides or actinides.

We find that there is not a significant impact on the \emph{total} lanthanide mass fraction ejected from each of the three disks; all three exhibit $\rm{X_{Lan}}$ values close to $3\times10^{-2}$. The mass of lanthanides ejected from each disk was more heavily influenced by the disk's total ejecta mass, rather than the ability of each disk to actually produce lanthanides, i.e. the lanthanide mass fraction.

In contrast, the magnetic field strength has a strong impact on actinide production. As listed in Table \ref{tab:masses}, increasing the initial field strength by a factor of ten, i.e. from \bten{} to \bcent, results in more than a factor of six larger actinide mass fraction. 

\subsection{Angular Dependence}

\begin{figure}
\centering
\includegraphics[scale=0.36]{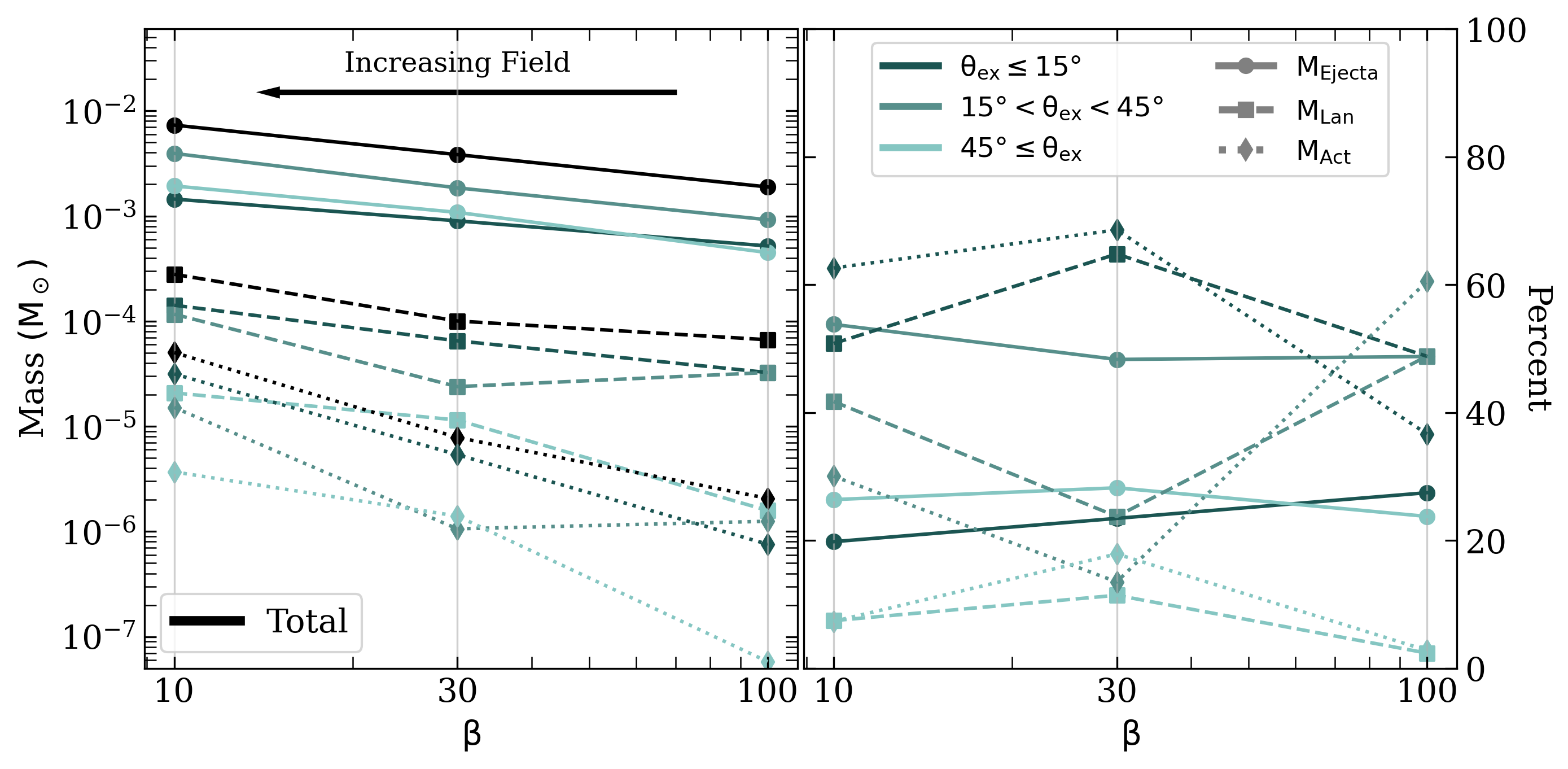}
\caption{\emph{Left:} Ejecta mass (solid), lanthanide mass (dashed), and actinide mass (dotted) produced in each of the angular components. \emph{Right:} Percent values of the contribution to the total ejecta mass (solid), the total lanthanide mass (dashed), and the total actinide mass (dotted) for the initial value of $\beta$ characterizing each disk, for each angular component. The light, medium, and dark blue lines show the contribution from the polar, intermediate, and equatorial outflow, respectively.}
\label{fig:percs}
\end{figure}

One of the most striking features of Figure \ref{fig:abs_angle} are the differing contributions to the overall \rp{} patterns from each angular part of the outflow from the three different disks. We plot the percentages of the total ejecta mass, total lanthanide mass, and total actinide mass originating from each angular cut in Figure \ref{fig:percs}. We also list these values in bottom half of Table \ref{tab:masses} for convenience. 

As can be seen by the solid light blue line on the left side of Figure \ref{fig:percs}, the largest total mass component was the one ejected at intermediate angles for all three disks; in each case, it accounted for roughly half the ejecta by mass. The other half of the mass was roughly evenly split between the polar and equatorial ejecta, with the \bten{} and \bthirty{} disks showing slightly more polar outflow and the \bcent{} disk showing slightly more equatorial outflow.

\subsubsection{Lanthanide Outflows}

\begin{figure*}[ht]
\centering
\includegraphics[scale=0.35]{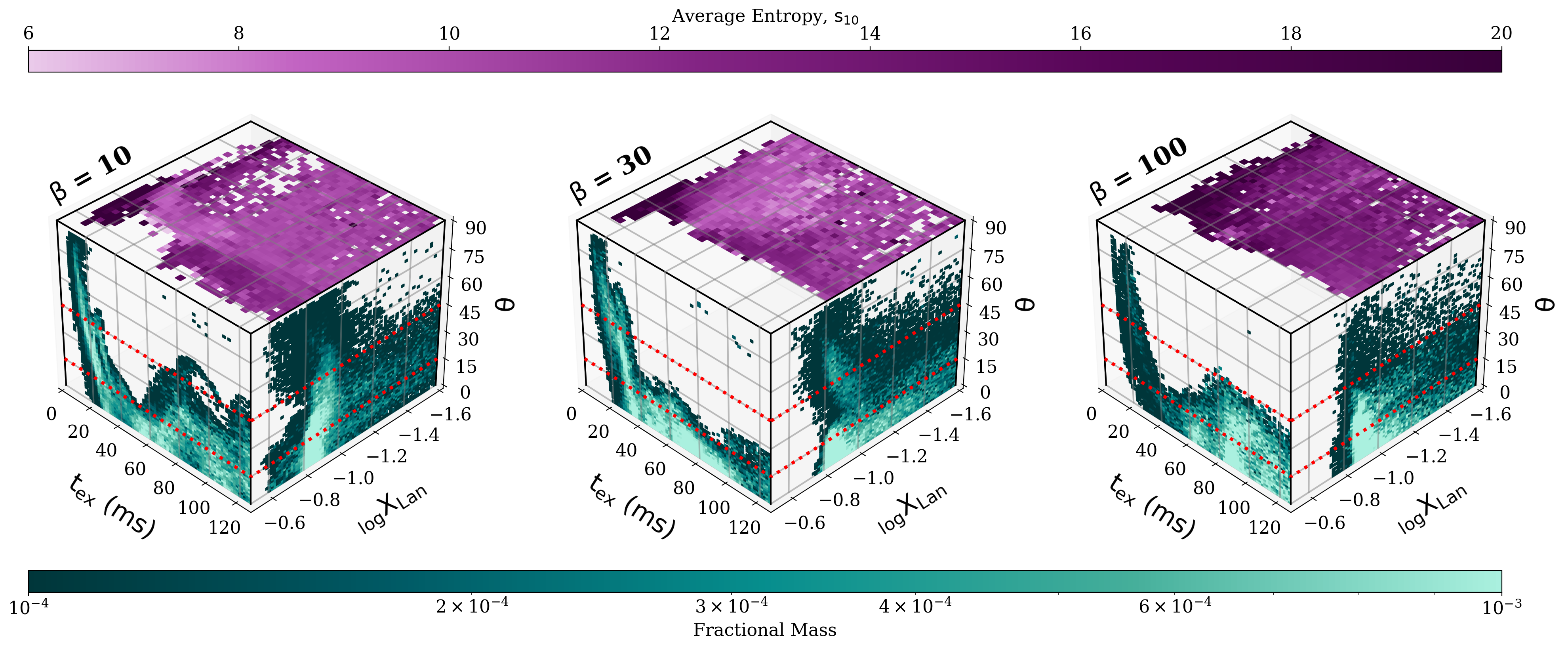}
\caption{\emph{Teal: } Two dimensional histograms showing the fractional mass binned by \textract{} (x-axis, left) or $\rm{log_{10}(X_{Lan})}$ (y-axis, right) and angle above the mid-plane (vertical axis). The color map shows the fraction of the total mass, i.e. all values sum to one on each face.
\emph{Purple: } Mapping of the average entropy as a function of \textract{} and $\rm{X_{Lan}}$. The disks are shown with increasing $\beta$ from left to right.}
\label{fig:histcube}
\end{figure*}

Since lanthanides are a critical component of opacity as well as stellar abundance pattern, we now turn from \textit{total} mass outflow to \textit{lanthanide} mass outflow. The dashed dark blue line on the left side of Figure \ref{fig:percs} indicates that the largest lanthanide mass emerges from equatorial angles, and this mass increases with increasing field strength. This figure also shows that the lanthanide mass ejected from polar angles contributes the least to the total lanthanide mass in all three disks, contributing less than or around 10\% of the total lanthanide mass in the \bten{} and \bcent{} disks.  

The intermediate angle component produces a lanthanide mass that is intermediate to the polar and equatorial lanthanide ejecta in all disks.  However it does not follow the same pattern of lanthanide mass increasing monotonically with increasing magnetic field strength. In both the \bten{} and \bcent{} disks, the lanthanide ejection from intermediate angles is close to comparable to that from the equatorial region.  However, in the \bthirty{} disk the intermediate angle lanthanides are only 24\% of the total lanthanide mass, compared to 40-50\% in the other two disks. This can be seen by comparing the medium blue squares (solid lines) in the right panel of Figure \ref{fig:percs}. The decreased lanthanide ejecta from intermediate angles in the \bthirty{} disk causes the overall lanthanide mass from the \bthirty{} to be the smallest of all three disks. While difficult to see on the log scale of the second panel of Figure \ref{fig:abs_angle}, the total lanthanide mass fraction in the \bthirty{} disk is almost one and a half times smaller than the other two disks.

At first glance, it would appear that these patterns reveal a simple decrease in lanthanide production with increasing angle. Indeed, this is true: the total lanthanide mass fraction in the equatorial ejecta is largest, followed by the intermediate angle ejecta, and the smallest lanthanide mass fraction coming from the polar ejecta. However, this does not tell the whole story.

Figure \ref{fig:histcube} shows the relation of \textract, $\rm{\theta_{ex}}$, lanthanide mass fraction, and average entropy in the form of a cube with shared axes. The teal color maps are two-dimensional mass histograms showing the fractional mass in each bin. The purple surface shows the same bins but colored by average entropy of the material in those bins. We included dotted red lines at values of $\theta_{ex}=15\degree$ and $\theta_{ex}=45\degree$ as a guide. Furthermore, we note that only tracers which resulted in $-1.6 < \rm{log_{10}(X_{Lan})} < -0.5$ are included in this sample.  

In all three disks, there is a transient-type component emerging early in time, at high angles. This initial transient is composed of on-average high entropy material, as shown by the dark values on the purple surface corresponding to an early time slice. The majority of the mass that emerges at high angles is ejected in this early ejection event, though the \bten{} disk shows subsequent mass ejection around \textract $=80$ ms. By examining the same high-angle slice on the sides showing $\rm{log_{10}(X_{Lan})}$, one can see that this material tends to result in fairly large lanthanide mass fraction, 
indicating that despite the overall higher \ye{} of the polar ejecta shown in Fig.  \ref{fig:basics_angle} some of the higher entropy polar ejecta is very effective at producing lanthanides.

Additionally, all three disks show equatorial ejection consistent with that shown in Figure \ref{fig:massoutflow}, indicating that the equatorially-ejected material fairly consistently produces lanthanides. This is further supported by the right faces of Figure \ref{fig:histcube}, which show the majority of the mass concentrated between $\rm{log_{10}(X_{Lan})}$ values of $-0.8$ and $-1.0$. 

As noted earlier in this section, the most variable lanthanide ejection in each of the disks is that which occurs at intermediate angles, and as can be seen from Figure \ref{fig:histcube}, these intermediate angles capture pulsations in the ejecta. Following the initial transient, the ejection angle of all three disks tends towards lower values. At some point following this, each disk shows another excess starting at around 50 ms. In the \bten{} disk, this second excess sees lanthanides getting ejected at angles as high as $60\degree$. After this excess, the beginning of another appears around 110 ms and again sees trace amounts of lanthanides being ejected above $45\degree$. The \bcent{} lanthanide outflow also shows material being pushed up to high angles, though not as high as the \bten{} case. The periodicity of these outflows roughly tracks the Keplerian orbital period of horizon-scale material in the disk, perhaps indicating a connection to low wave-number oscillation modes in the disk \citep{Abramowicz2006}, such as those caused by the magnetic dynamo \citep{Heinemann2009}.
\footnote{For a review of these phenomena, see \citet{Abramowicz2013}.}

Once again, the \bthirty{} behaves rather differently. Following the initial transient event, lanthanides are only ejected around or below $30\degree$ above the mid-plane. Furthermore, beyond around 100 ms, only fairly small amounts of lanthanides are ejected from any angle. This leads to the previously noted lower overall abundance of lanthanides coming from this disk.

\subsubsection{Actinide Outflows}

While increased lanthanide mass ejected in disks with stronger magnetic fields is more a consequence of larger ejecta masses overall, this is not the case for the actinides: in addition to disks with stronger fields ejecting more actinide mass, one can see from Table \ref{tab:masses} that increasing the magnetic field tenfold results in more than a factor of six larger actinide mass fraction. This is also apparent from the scaled total abundance patterns shown in the top panel of Figure \ref{fig:abs_angle}, where the third peak and actinide abundances show some of the largest variability. 

The largest contribution to the actinide outflow at higher magnetic field comes from the equatorial region as one might expect, but as can be seen in the left panel of Figure \ref{fig:percs}, the largest contribution to the total actinide mass in the lowest magnetic field disk comes not from the equatorial region but from the intermediate angle ejecta.  However, in this \bcent{} disk, the vast majority of the remaining actinide production does come from the equatorial outflow. In \bthirty{}, the sub-dominant ejecta is divided reasonably similarly between the intermediate and polar ejecta, while in the \bten{} disk the polar actinide outflows play only a small role. These angular patterns indicate a much higher variability in actinide production depending on magnetic field strength as compared to lanthanide production.

\section{Conclusion and Outlook}\label{sec:conclusion}

We performed three separate GR$\nu$MHD simulations representing a NSM black-hole accretion disk. For each disk, we varied the parameter $\beta$ as a proxy for varying initial magnetic field strength. We followed the evolution of each disk out to $10^4\ \rm{GM_{BH}c^{-3}}$, or roughly 127 ms. We extracted more than $400000$ Lagrangian tracer particles from these simulations to compute nucleosynthetic yields.

We found that increasing the initial magnetic field strength by a factor of ten resulted in the disk ejecting half its mass more than 10 ms earlier. Furthermore, doing so resulted in almost a factor of 4 larger total ejecta mass. In each case, roughly half the material was ejected between $15\degree$ and $45\degree$.

We performed nucleosynthesis on each of the unbound tracer particles and found broad agreement: all three disks produced a full \rp{} pattern including actinides. In each case, the total lanthanide mass fraction was between $2-4\times10^{-2}$ ; the total lanthanide mass roughly scaled as the total ejecta mass. The smallest contribution to the lanthanides came from polar material ejected higher than $45\degree$ above the mid-plane.

We found large variation in the actinide production as a function of magnetic field strength. In the stronger field cases of the \bten{} and \bthirty{} disks, more than half of the actinide content produced originated from equatorial ejecta. Meanwhile, the largest percentage of actinide mass in the \bcent{} disk emerged from the intermediate angle ejecta. We found that decreasing $\beta$ from 100 to 10 resulted in more than a factor of 6 larger total actinide mass fraction. 

Our observation that changing the initial magnetic field strength in the three disks resulted in roughly similar lanthanide mass fractions but very different actinide mass fractions hints at a possible connection to the wide variability in actinide enhancements exhibited in stellar observations of \rp-enhanced, metal-poor stars (see, for example, \citet{Holmbeck2018,Holmbeck2020,Lund2023,Kullmann2023} for recent studies addressing this question\footnote{See section 4.2 of \citet{Holmbeck2023_rev} for a review.}). 

Differences in the evolution of the isotopic composition of the ejecta (and therefore the nuclear heating) can imprint distinct signatures in kilonova light curve predictions caused by specific nuclei \citep{Zhu2018,Vassh2019,Even2020,Zhu2021,Barnes2021,Lund2023,Kedia2023}. We therefore expect that recording the nuclear heating profile based on the angle dependence of the ejecta we have found in this work would yield similar consequences for angle-dependent kilonova predictions. Due to the computational cost of running such a large number of nucleosynthesis calculations, we extracted the abundances at a (single) late time in order to investigate broadly what abundance patterns these systems could produce. A complete picture of the impact of magnetic fields on kilonova observables requires passing nucleosynthetic yields and outflow morphology through a full radiative transfer calculation, where we expect as rich a phenomenology in the light curves and spectra as we have found in the nucleosynthesis.

\section{Acknowledgements}
This work is approved for unlimited release with LA-UR-23-32340.
We gratefully acknowledge the support of the U.S. Department of Energy through the Laboratory Directed Research and Development (LDRD) program and the Center for Nonlinear Studies (CNLS) at Los Alamos National Laboratory for this work. 
This research also used resources provided by the LANL Institutional Computing Program.
LANL is operated by Triad National Security, LLC, for the National Nuclear Security Administration of U.S. Department of Energy (Contract No. 89233218CNA000001). 
This work was partially supported by the Fission in r-Process Elements (FIRE) topical collaboration in nuclear theory, funded by the U.S. DOE, contract No. DE-AC5207NA27344.
K.A.L. acknowledges support from the Seaborg Institute for funding under LDRD project 20210527CR.
J.M.M. acknowledges support from LDRD project 20220564ECR. 
M.R.M. acknowledges support from LDRD project 0230052ER. 
We acknowledge support from the NSF (N3AS PFC) grant No. PHY-2020275, as well as from U.S. DOE contract Nos. DE-FG0202ER41216 and DE-SC00268442 (ENAF), as well as by LA22-ML-DE-FOA-2440.

\bibliography{ref}

\end{document}